\documentclass[12pt]{iopart}
\usepackage[english]{babel}
\usepackage[utf8]{inputenc}
\usepackage{pgf,tikz,pgfplots}
\pgfplotsset{compat=1.15}
\usetikzlibrary{arrows}
\usepackage[utf8]{inputenc}
\usepackage{amssymb}
\usepackage{bm}
\usepackage{csquotes}
\expandafter\let\csname equation*\endcsname\relax
\expandafter\let\csname endequation*\endcsname\relax
\usepackage{appendix}
\usepackage{array}
\usepackage{amsmath}
\usepackage{amsfonts}
\usepackage{amstext}
\usepackage{amssymb}
\usepackage{physics}
\usepackage{nicematrix}
\usepackage{amsmath}
\usepackage{amsfonts}
\usepackage{amssymb}
\usepackage{physics}
\usepackage{dsfont}
\usepackage{color}
\usepackage{hyperref}
\usepackage{inputenc}
\usepackage{enumerate}
\usepackage{graphicx}
\usepackage{float}
\usepackage{amsmath}
\usepackage{amsfonts}
\usepackage{setspace} 
\usepackage{lipsum}
\usepackage{dcolumn}
\usepackage{hyperref}
\usepackage{subfigure}
\usepackage{epsfig}
\usepackage{epsf}
\usepackage{multirow}
\usepackage{epstopdf}

\usepackage{color}
\usepackage{enumitem}
\usepackage{bm}
\usepackage{subfigure}
\usepackage{titlesec}
\usepackage{tabularx}
\usepackage{xcolor}
\usepackage{tikz}
\usepackage[margin=1in]{geometry}
\usepackage{pgfplots}
\usepackage[justification=justified, margin=0cm]{caption}

\begin{document}

\title[Magnetic field localization effects on heat transport in a harmonic chain ]{Localization effects due to a random magnetic field on  heat transport in a harmonic chain}

\author{Ga\"etan Cane$^1$, Junaid Majeed Bhat$^2$, Abhishek Dhar$^3$ and C\'edric Bernardin$^{4}$}

\bigskip
\address{${}^{1}$ Universit\'e C\^ote d'Azur, CNRS, LJAD, Parc Valrose, 06108 NICE Cedex 02, France}\ead{Gaetan.cane@univ-cotedazur.fr}

\bigskip

\address{${}^{2}$ International Centre for Theoretical Sciences, Bengaluru, India-560089}\ead{junaid.bhat@icts.res.in}
\bigskip

\address{${}^{3}$ International Centre for Theoretical Sciences, Bengaluru, India-560089}
\ead{abhishek.dhar@icts.res.in}

\bigskip
\address{${}^4$ Universit\'e C\^ote d'Azur, CNRS, LJAD, Parc Valrose, 06108 NICE Cedex 02, France \\
Interdisciplinary Scientific Center Poncelet (CNRS IRL 2615), 119002 Moscow, Russia}
\ead{cbernard@unice.fr}


\begin{abstract}
We consider a  harmonic chain of $N$ oscillators in the presence of a disordered magnetic field. The ends of the chain are connected to heat baths and we study the effects of the magnetic field randomness on  heat transport. The disorder, in general, causes localization of the normal modes due to which a system becomes insulating. However, for this system, the localization length diverges as the normal mode frequency approaches zero. Therefore, the low frequency modes contribute to the transmission, $\mathcal{T}_N(\omega)$, and the heat current goes down as a power law with the system size, $N$. This power law is determined by the small frequency behaviour of some Lyapunov exponent, $\lambda(\omega)$, and the transmission in the thermodynamic limit, $\mathcal{T}_\infty(\omega)$. While it is known that in the presence of a constant magnetic field $\mathcal{T}_\infty(\omega)\sim \omega^{3/2},~ \omega^{1/2}$  depending on the boundary conditions, we find that the Lyapunov exponent for the system behaves as $\lambda(\omega)\sim\omega$ for $\expval{B}\neq 0$ and   $\lambda(\omega)\sim\omega^{2/3}$ for $\expval{B}=0$. Therefore, we obtain different power laws for current vs $N$ depending on $\expval{B}$ and the boundary conditions.   
\end{abstract}

\noindent{\it Keywords\/}: Heat conduction, Transport properties.

%
%
%
%
%

\section{Introduction}

In his seminal paper \cite{Anderson58} Anderson studied the conductance of electrons and explained how the presence of impurities in the metal could reduce drastically the diffusive motion of the electrons up to a complete halt and thus giving place to an insulator. This phenomenon depends strongly on the dimension and the Metal Insulator Transition is not instantaneous with respect to the disorder strength only in dimension greater or equal to three.  Nowadays Anderson localization is seen as a generic phenomenon present in disordered media whereby the addition of random defects in the medium has the tendency to localize in space the normal modes of the system. As a consequence it reduces drastically the transport coefficient. Originally Anderson's work took place in a quantum context but the phenomenon he explained appears also in a classical one. In the 70's Lebowitz and others \cite{CasherLebowitz71, RubinGreer71, OConnorLebowitz74,Verheggen79} started to investigate the effect of impurities (random masses) on the transport properties of a one dimensional harmonic chain, arguing in particular that the conductivity $\kappa (N)$ of the chain (which is proportional to the system size $N$ for a purely harmonic chain), loses some order of magnitude because of disorder: $\kappa (N)\sim N^a$, $a<1$ (see \cite{AjankiHuveneers11} for a mathematical proof). But at the difference with respect to original Anderson localization, the conductivity does not become exponentially small in the system size and, depending on the physical boundary conditions and thermostats, it can vanish ($a<0$), diverge ($0<a<1$) or even converge ($a=0$) \cite{Dhar01,RoyDhar08, DeRoeckDharHuveneersSchutz17}.  The reason for this is roughly due to the fact that for disordered harmonic chains, normal modes with frequency $\omega$ becomes localized but with a length of localization $\ell (\omega) \sim \omega^{-2}$. The role of thermostats and boundary conditions is more difficult to explain without going into computational details. Thus we note that in disordered harmonic chains the  low frequency modes have still the possibility to transport energy \cite{BernardinHuveneersOlla19}. The case where the disorder is in the interparticle springs instead of the masses has recently been addressed in ~\cite{amir2018,ash2020}.
In higher dimensions the situation is less understood \cite{LeeDhar05,Chauduryetal10}. 
More recently there has been a renewed interest for these questions with respect to the effect of nonlinearities \cite{DharLebowitz08, DharSaito08} or of an energy conserving noise \cite{Bernardin08, DharVenkateshenLebowitz11, BernardinHuveneers13, BernardinHuveneersLebowitzLiverani15}.\\

In this paper we consider an ordered (constant masses) one-dimensional chain of two-dimensional charged oscillators subject to a  random transverse magnetic field on every lattice site (or equivalently a chain of disorderly charged oscillators  subjected to a constant magnetic field on the lattice). In \cite{ordered2021}, by using the Non-Equlibrium Greens's Function (NEGF) formalism we obtained an explicit expression of the heat current in the steady state and investigated then the transport properties of such a system when all the charges are the same. We established that transport is ballistic like for ordered harmonic chains. The aim of this paper is therefore to describe the effect of the charge impurities on the behaviour of the conductivity of the system. This will require us to investigate the frequency dependence of the localization lengths of normal modes. We show that due to charge disorder the current shows different scaling with the system size which depends on the boundary conditions as well as on the expectation value of the magnetic field.\\

This paper is structured as follows: in Sec.~\ref{sec:model} we introduce the model and state the results for  heat current  using the NEGF formalism. We also present numerical results for the transmission function and discuss the effects of localization due to the random magnetic field on  the transmission function. In Sec.~\ref{sec:lyapunov}, we use  the Green's function expression as a product of random matrices to determine the Lyapunov exponents.  We also present numerical results for Lyapunov exponents which are consistent with our theoretical results. Using the results for the Lyapunov exponents, we finally determine the size dependence of the mean of the heat current in Sec.~\ref{sec:size_current} and compare with direct numerical calculations of the current. We conclude in Sec.~\ref{sec:concl}.

\section{The Model and  heat current by NEGF}
\label{sec:model}
\subsection{The model}
We consider a  chain of $N$  harmonic oscillators each having two transverse degree's of freedom so that every oscillator is free to move in a plane perpendicular to the length of the chain. We choose the plane of motion to be the $x-y$ plane and  denote  the positions and momenta of the  $n^{th}$ oscillator by $(x_n,y_n)$ and $(p_{n}^x,p_{n}^y)$ respectively, with $n=1,2,\ldots,N$. The oscillators are assumed to have unit masses  and each carry a positive unit charge. We consider a magnetic field $\vec B_n=B_n {\vec{\mathbf e}}_z$ perpendicular to the plane of motion which can be obtained from a vector potential $\vec {\bf A}_n=(-B_n y_n,B_n x_n,0)$ at each lattice site. In this paper, we assume that $(B_n)_n$ form a sequence of independent identically distributed random variables with average $\langle B \rangle$ and variance $\sigma^2$. The Hamiltonian of the chain is given by:
\begin{align*}
H=\sum_{n=1}^N \frac{(p_{n}^x+B_n y_n)^2  +(p_{n}^y- B_n x_n)^2}{2}&+ \sum_{n=0}^N \frac{(x_{n+1}-x_n)^2 + (y_{n+1}-y_n)^2}{2},
\end{align*} 
where the inter particle spring constant has been fixed to $1$. We will consider the two different boundary conditions: (i) fixed boundaries with $x_0=x_{N+1}=0$ and (ii) free boundaries with $x_0=x_1,~x_N=x_{N+1}$. In order to study heat current through this system, we consider the $1^{\rm st}$ and the $N^{\rm th}$ oscillators to be connected to heat reservoirs at temperatures $T_L$ and $T_R$ respectively. The heat reservoirs are modelled using dissipative and noise terms leading to the  following Langevin equations of motion:
\begin{align}	
\ddot x_n&=(x_{n+1}+x_{n-1}-c_n x_{n})+  B_n \dot y_n+\eta_L^x(t)\delta_{n,1}+\eta_R^x(t)\delta_{n,N}-(\gamma \delta_{n,1}+\gamma\delta_{n,N})\dot x_n \label{xiddot}\ ,\\
\ddot y_n&= (y_{n+1}+y_{n-1}-c_n y_{n})-  B_n \dot x_n +\eta_L^y(t)\delta_{n,1}+\eta_R^y(t)\delta_{n,N}-(\gamma \delta_{n,1}+\gamma\delta_{n,N})\dot y_n \label{yiddot}\ .
\end{align}
for $n=1,2,\ldots,N$. Here $\eta_L (t) :=(\eta_L^x(t),\eta_L^y(t))$ and  $\eta_R (t):=(\eta_R^x(t),\eta_R^y(t))$ are Gaussian white noise terms acting on the  $1^{\rm st}$ and $N^{\rm th}$ oscillators respectively. These follow the regular white noise correlations, $\expval{\eta_{L/R}(t)\eta_{L/R}(t')}=\sqrt{2\gamma T_{L/R}}\delta(t-t')$ (Boltzmann's constant is fixed to one to simplify), where $\gamma$   is the dissipation strength at the reservoirs.   The coefficients $c_n$ fix the boundary conditions of the problem. For fixed boundaries $c_n=2$ for all $n$, while for free boundary conditions $c_n=2-\delta_{n,1}-\delta_{n,N}$.

\subsection{Heat current}

In \cite{ordered2021}, by using the non-equilibrium Green function formalism, we obtained an exact expression  for the heat current, ${J}_N$, in the  steady state of the chain. More exactly, let us define the processes   $(f_n^\pm)_{n\ge 0}$ and $(g_n^\pm)_{n\ge 0}$ as
  \begin{equation}
    \label{f_N}
  \begin{split}
 f_{n+1}^\pm&=(c_{n+1} -\omega^2 \pm \omega B_{n+1}) f_n^\pm-f_{n-1}^\pm, \quad  f_0^\pm =1, \quad f_1^\pm =c_1-\omega^2\pm\omega B_1 \ ,\\
 g_{n+1}^\pm&=(c_{n+1} -\omega^2 \pm \omega B_{n+1}) g_{n}^\pm-g_{n-1}^\pm, \quad g_0^\pm =0, \quad g_1^\pm =1 \ .
\end{split}
\end{equation}
Then we introduce 
\begin{equation}
 \label{eq:FN}
  F_N^\pm :=F_N^\pm (\omega) =  f_N^\pm+i {\gamma} \omega (g_{N}^\pm+f_{N-1}^\pm)-{\gamma^2}  \omega^2 g_{N-1}^\pm \ , 
\end{equation}
and the  heat current is equal to 
   \begin{equation}
   \label{eq:exprCurrent}
{J}_N =(T_L-T_R) \, \int_{-\infty}^{\infty} d\omega \, {\mathcal T}_N (\omega) = 2 (T_L-T_R) \, \int_{0}^{\infty} d\omega \, {\mathcal T}_N (\omega) 
  \end{equation}
with the net transmission function ${\mathcal T}_N$ defined for any frequency $\omega$ by
\begin{equation}
{\mathcal T}_N (\omega) := \frac{\gamma^2}{\pi} \ \omega^2 \ \left[\frac{1}{\abs{F_N^+ (\omega)}^2}+\frac{1}{\abs{F_N^- (\omega) }^2}\right]. \label{taueqn2}
\end{equation}

We denote by $\langle {J}_N \rangle$ the expectation of the heat current with respect to the magnetic field distribution $\langle \cdot \rangle$ and our goal is to understand its scaling behavior in $N$.\\

Observe that the stochastic processes $(f_n^-)_{n\ge 0}$ and $(g_n^-)_{n\ge 0}$ are defined in terms of the two dimensional discrete time Markov chain $(U_n )_{n\ge 0}$  given by 
\begin{equation}
\label{eq:MarkovChain}
U_{n+1}= \begin{pmatrix} 2-\omega^2 - \omega B_{n+1}&-1 \\1&0  & \end{pmatrix} \ U_{n}, \quad \text{where}\quad  U_n:=\begin{pmatrix} u_{n}\\ u_{n-1}\end{pmatrix},
\end{equation}
by choosing suitable initial conditions. By replacing the $B_n$'s by $-B_n$'s in the last display, we see that $(f_n^+)_{n\ge 0}$ and $(g_n^+)_{n\ge 0}$ can also be expressed in terms of $(U_n)_{n\ge 0}$ . The state of the Markov chain is nothing but the result of a product of $2\cross 2$ product of independent and identically distributed random matrices.  Roughly, the behaviour of $F_N^\pm$ is related to the growth of $\Vert U_N (\omega) \Vert$ which will be in the form $e^{2 \lambda (\omega) N}$, where 
\begin{equation}
\label{eq:lyapunov00}
\lambda (\omega) = \lim_{n \to \infty} \cfrac{1}{2n} \left\langle \log \Vert U_n (\omega)\Vert \right\rangle =\lim_{n\to \infty} \cfrac{1}{n} \left\langle \log \vert u_n (\omega)\vert \right\rangle >0,
\end{equation}
with $\left\langle...\right\rangle$ denoting a disorder average, is half the Lyapunov exponent associated to the Markov chain  $(U_n)_{n\ge 0}$, or equivalently of the corresponding product of random matrices. The limit exists by Furstenberg's Theorem \cite{Furstenberg63}, is non-negative, independent of the initial condition $U_0$ and the limit holds in fact also for any realisation and not only by averaging over the magnetic field distribution. 

For now we quickly discuss the effect of localization due to the random magnetic field on the heat transport and the need for calculating the Lyapunov exponent $\lambda (\omega)$ for small frequencies $\omega$.
\subsection{Effect of localization due to random magnetic field on  the  net Transmission}
\begin{figure}[t!]
	\centering
	\subfigure[Uniform magnetic field]{
		\includegraphics[width=7.65cm,height=7.75cm]{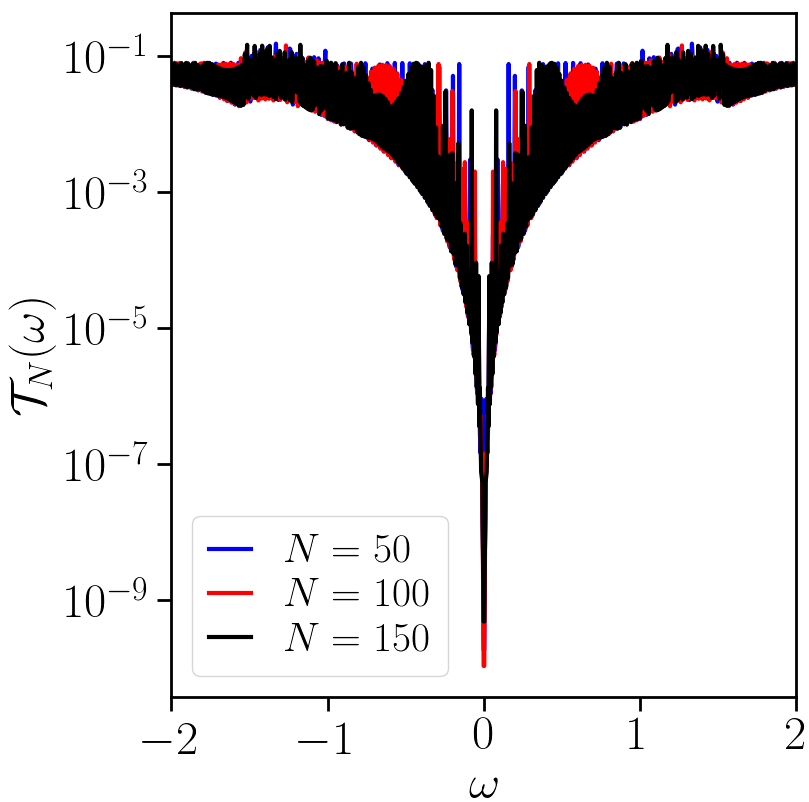}
	}
	\subfigure[Random magnetic field]{
		\includegraphics[width=7.65cm,height=7.65cm]{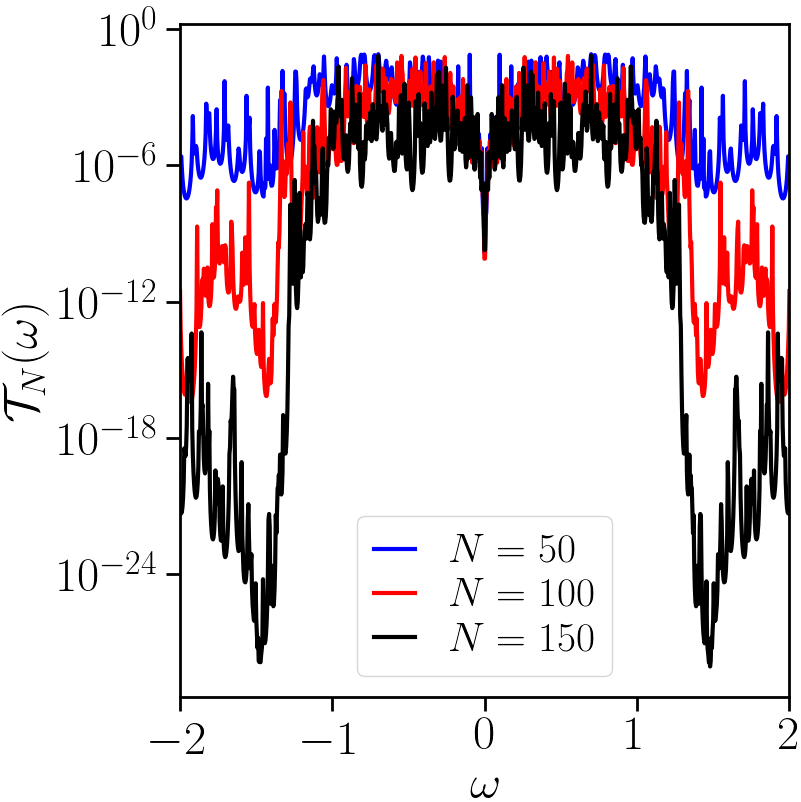}
	}	
	\caption{Variation of the net transmission, in units of $k_B=1$, with $\omega$ for uniform magnetic field, panel (a), and random magnetic field, panel (b). The axes are in log scale and $\gamma=0.2$. The magnetic field in (a) is set to be $1$ on all oscillators and  in (b) it was chosen uniformly from the interval $(0,2)$. As can be seen clearly from the plots, the localization effects cause suppression of the transmission.}\label{tau_plts}
\end{figure}
 Using Eq.~\eqref{taueqn2}, we can calculate the net transmission $\mathcal{T}_N(\omega)$ for any spatial configuration of the magnetic field  using a computer programme. In Fig.~(\ref{tau_plts}a) and Fig.~(\ref{tau_plts}b) we plot the net transmission function with $\omega$ for a uniform magnetic field and for a random magnetic field  for different system sizes respectively. On comparison of the two plots, we can see that the randomness causes suppression of the net transmission and also the net transmission for the random magnetic field case goes down with system size while the system size has nearly no effect on  the transmission for the uniform magnetic field. The suppression in case of random magnetic field is due to localization of the normal modes of the system. The normal modes of frequency $\omega$ get exponentially localized due to randomness with a localization length given by $1/ \lambda(\omega)$ where $\lambda(\omega)$ is the Lyapunov exponent defined in Eq.~\eqref{eq:lyapunov00}. As a result of this they a priori do not contribute to the transmission. However, note that the transmission for random magnetic field is higher near $\omega=0$ and goes down as we move away which means that the normal modes with energies closer to $\omega=0$ have a  larger localization length, i.e. $\lambda (\omega) \to 0$ as $\omega \to 0$. Since we are eventually interested in the size dependence of the current, for large $N$, which is the integral of the transmission over all $\omega$, we can reduce the integration limit to values of $\omega$ for which the localization length is greater than the system size. For the remaining $\omega$ values for which the localization length is less than the system size, the transmission would be negligible. Hence, we cut off the integral limit to $\omega=\omega^N_{max}$ where  $1/ \lambda(\omega^N_{max})=N$ and the current is then given by
  \begin{equation}
\left\langle{J}_N \right\rangle \approx 2(T_L-T_R)\int_0^{\omega^N_{max}} d\omega  \lim\limits_{N\rightarrow\infty} \left\langle \mathcal{T}_N(\omega)\right\rangle=2(T_L-T_R)\int_0^{\omega^N_{max}} d\omega   \mathcal{T}_\infty(\omega) \ .\label{currexp1}
 \end{equation}
 Note that the frequency $\omega^N_{max}$ would be very small for large $N$, and for such small frequencies we expect  $\mathcal{T}_\infty(\omega)$  to have a weak dependence on disorder [since in the recursion  Eq.~\eqref{eq:MarkovChain}, the randomness is multiplied by $\omega$] --- hence in the above equation  $\mathcal{T}_\infty(\omega)$ is written without a disorder average and can in fact be determined by considering the chain in a \emph{constant} magnetic field of strength  $\expval{B}$.   In \cite{ordered2021}, we proved that for constant magnetic field $\expval{B}\neq 0$, $\mathcal{T}_\infty(\omega)\sim \omega^{3/2}$ and $\sim \omega^{1/2}$  for fixed and free boundaries respectively, while for $\expval{B}=0$ it goes as $\omega^2$ and $\omega^0$ for the two boundary conditions respectively.  To determine the size dependence of the current in addition to the small $\omega$ behaviour of $\mathcal{T}_\infty(\omega)$ we also need the small $\omega$ behaviour of $\lambda(\omega)$. We now proceed to the next section where we discuss  the  Lyapunov exponents of this equation.

\section{Analysis of the Lyapunov exponents}
\label{sec:lyapunov}
In this section we present theoretical and numerical results on the asymptotics of  Lyapunov exponents for small $\omega$ for the Markov processes defined by Eq.~\eqref{eq:MarkovChain}. The Lyapunov exponents are independent of the boundary conditions  -- so for this section we only work with fixed boundary conditions by setting $c_n=2$ for all $n$ -- and of the initial condition of the process -- i.e. it is the same for $f_n^\pm$ and $g_n^\pm$. We show that Eq.~\eqref{eq:MarkovChain} has three different behaviors for the Lyapunov exponent depending on the expected value $\langle B\rangle$ of the random magnetic field. For $\langle B \rangle>0$ the Lyapunov exponent satisfies $\lambda(\omega) \sim \omega$ and for $\langle B \rangle<0$,  $\lambda(\omega)\sim \omega^{1/2}$. However, for $\langle B \rangle=0$, $\lambda(\omega)\sim \omega^{2/3}$. Similar  Lyapunov exponent behaviours are found for a harmonic oscillator with parametric noise,~\cite{Crauel1999} and we will see that Eq.~\eqref{eq:MarkovChain}  could be written exactly in this form in the continuum limit.

\subsection{Theoretical results for Lyapunov exponents}
\label{sec_theoretical_result_lyapunov}
Let $(z_t)_{t \geq 0} \in {\mathbb R}^2$ be the solution of the following stochastic differential equation (with arbitrary initial condition)
\begin{equation}
\label{SDE_z}
\dot{z_t} = A_0 z_t  + \varepsilon \sigma \xi_t \, A_1 z_t \ ,
\end{equation}
where  $\varepsilon$ is a small positive parameter, $\sigma>0$ a constant, $\xi_t$ a one dimensional standard white noise and $A_0$ and $A_1$ are $2 \times 2$ matrices such that 
\[ A_0 = \begin{pmatrix}
0 & 1\\
-c & 0
\end{pmatrix},  \quad A_1 = \begin{pmatrix}
0 & 0\\
-1 & 0
\end{pmatrix}  \quad \text{with $c \in \mathbb{R}$}\ .\]
The Lyapunov exponent $\lambda_z(\varepsilon)$ of the process $(z_t)_{t \geq 0}$ is defined by
\begin{equation}
\label{Lyapunov_def_appendix}
\lambda_z (\varepsilon) = \lim_{t \rightarrow \infty} \frac{1}{t} \left\langle \log \Vert z_t \Vert \right\rangle ,
\end{equation}
where $\langle \cdot \rangle$ denotes the expectation with respect to the white noise. It is proved in \ref{sec:app-lyap} that  if we denote $z_t =(u_t, v_t)^\bot$ then we have the Lyapounov exponent for $(u_t)_{t\ge 0}$ is the same as for $(z_t)_{t\ge 0}$:
\begin{equation}
\label{eq:lambdaxz}
\lambda_z (\varepsilon)=\lim_{t \rightarrow \infty} \frac{1}{t} \left\langle \log \vert u_t \vert \right\rangle.
\end{equation}
The following result, proved in \cite{Wihstutz1999}, gives the behaviour of the Lyapunov exponent $\lambda_z (\varepsilon)$ for small noise
 \begin{enumerate}
 \item If $c=0$ then $\lambda_z(\varepsilon) = \hat{\lambda} (\sigma) \varepsilon^{2/3}$ where $\hat{\lambda} (\sigma)$ is defined in Eq. (\ref{expression_coef_<B>=0}) .
 \item If $c>0$ then $\lambda_z(\varepsilon) \sim \frac{\sigma^2}{8c} \varepsilon^2$ .
 \item If $c<0$ then $\lambda_z(\varepsilon) \sim \sqrt{\vert c \vert}$ . 
 \end{enumerate} 
A sketch of the proof of this result is given in \ref{sec:app-lyap}.\\

Consider now Eq.~\eqref{eq:MarkovChain} defining the discrete time Markov chain $U_n= (u_{n}, u_{n-1})^\top$ and rewrite it in the following form, for small $\omega$,
\begin{equation*}
u_{n+1} +u_{n-1} -2 u_n =- \omega \langle B\rangle  u_{n}  - \omega (B_{n+1}-\langle B\rangle) u_n +{\mathcal O} (\omega^2)\ .  
\end{equation*}
In the continuum limit, the discrete time process $(u_n)_{n\ge 0}$ becomes then the continuous time process $(u_t)_{t\ge 0}$ solution of
\begin{equation}
\label{eq:eqx}
\ddot{u}_t= - \omega \langle B \rangle  u_t -\omega \sigma  \xi_t u_t 
\end{equation}
where $(\xi_t)_{t\ge 0}$ is a standard white noise and $\sigma^2$ the variance of the $(B_n)_n$. Defining $w_t=(u_t, {\dot u}_t)^\top $ we see that the previous equation reads 
\begin{equation}
 \dot w_t = \begin{pmatrix}
0 & 1\\
- \omega \langle B \rangle & 0
\end{pmatrix} w_t + \sigma\omega \xi_t \begin{pmatrix}
0 & 0\\
-1 & 0
\end{pmatrix} w_t \ .
\label{zeqn1}
\end{equation}
We are interested in the Lyapunov exponent of the process $(u_t)_{t \geq 0}$ (or equivalently of the process $(w_t)_{t\ge 0}$ as said before):
\begin{equation}
\label{eq:lambda_w}
\lambda_w (\omega) =\lim_{t\to \infty} \cfrac{1}{t} \left\langle \log \Vert w_t\Vert \right\rangle = \lim_{t\to \infty} \cfrac{1}{t} \left\langle \log \vert u_t\vert \right\rangle
\end{equation}
\newline Eq.~\eqref{zeqn1} looks similar to Eq.~(\ref{SDE_z}) but to fit perfectly with Eq.~\eqref{SDE_z} we perform the time scaling
\begin{equation*}
{\tilde u}_t= u_{ t /\sqrt \omega}
\end{equation*}
in Eq.~\eqref{eq:eqx} wich gives by scaling invariance of white noise
\begin{equation}
\label{eq:eqx_scaled}
\ddot{\tilde u}_t= - \langle B \rangle  {\tilde u}_t -\omega^{1/4}  \sigma\xi_t {\tilde u}_t
\end{equation}
or equivalently for ${\tilde z}_t =({\tilde u}_t, \dot{\tilde u}_t)^\top$ the equation
\begin{equation}
 \dot {\tilde z}_t = \begin{pmatrix}
0 & 1\\
- \langle B \rangle & 0
\end{pmatrix} {\tilde z}_t + \sigma\omega^{1/4} \xi_t \begin{pmatrix}
0 & 0\\
-1 & 0
\end{pmatrix} {\tilde z}_t \ .
\label{zeqn_scaled}
\end{equation}
With the previous notation we have hence
\begin{equation}
\label{link_between_lyapunov_exponent_scaled}
\lambda_w (\omega) = \sqrt{\omega} \ {\lambda}_{\tilde z} (\omega^{1/4})\ .
\end{equation}
Eq.~\eqref{zeqn_scaled} fits perfectly Eq.~\eqref{SDE_z} with $c= \langle B \rangle$ and $\varepsilon = \omega^{1/4}$. Then using point (i), (ii) and (iii) of Eq.~\eqref{SDE_z} and Eq.~\eqref{link_between_lyapunov_exponent_scaled} we get
\begin{enumerate}
\item If $\langle B \rangle = 0$, $\lambda_w (\omega) = \hat{\lambda} (\sigma)  \omega^{2/3}$ where $\hat{\lambda} (\sigma)$ is defined in Eq. (\ref{expression_coef_<B>=0}) \ .
\item  If $\langle B \rangle > 0$, $\lambda_w (\omega)\sim \frac{\sigma^2}{8\expval{B}}\omega$\ .
\item If $\langle B \rangle < 0$, $ \lambda_w (\omega)\sim \sqrt{\vert \expval{B} \vert }\omega^{1/2}$\ .
\end{enumerate}

\medskip

It makes sense to believe that $\lambda (\omega)$ defined by Eq.~\eqref{eq:lyapunov00} and $\lambda_w (\omega)$ defined by Eq.~\eqref{eq:lambda_w} have roughly the same behaviour as $\omega \to 0$ but a strong theoretical argument supporting this belief is missing. However, in the case $\langle B \rangle >0$, we can obtain directly the behaviour of $\lambda (\omega)$ by following the approach of \cite{MatsudaIshii70} and we observe then a good agreement at first order between $\lambda (\omega)$ and $\lambda_w (\omega)$, not only at the level of the exponent in $\omega$ but also at the level of the prefactor, see Table \ref{table:prefactorandexponent}. Unfortunately we were not able to carry this approach for $\langle B \rangle <0$ or $\langle B \rangle =0$ and we decided hence to not pursue this approach. However numerical results presented in the next section support strongly the claim that $\lambda(\omega) \sim \lambda_w (\omega)$ for $\omega \to 0$.

\subsection{Numerical results for Lyapunov exponents}

\begin{figure}
\centering
\subfigure[$\expval{B}>0$]{
	\includegraphics[width=5.1cm,height=5.1cm]{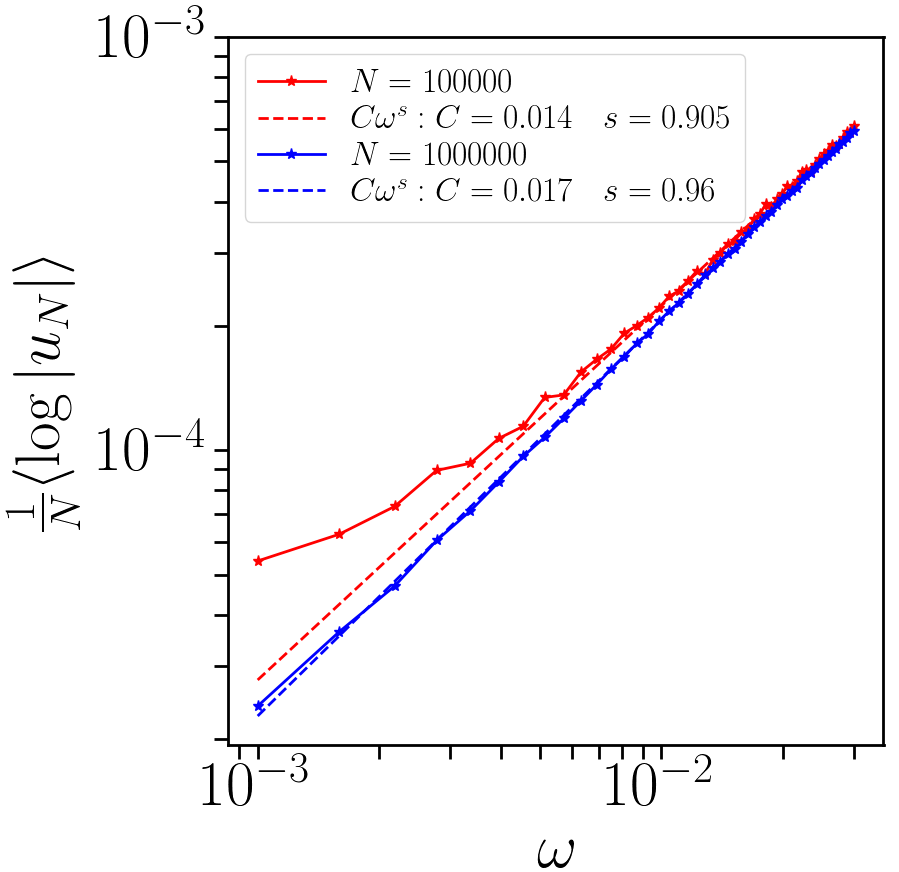}
}
\subfigure[$\expval{B}=0$]{
	\includegraphics[width=5.1cm,height=5.1cm]{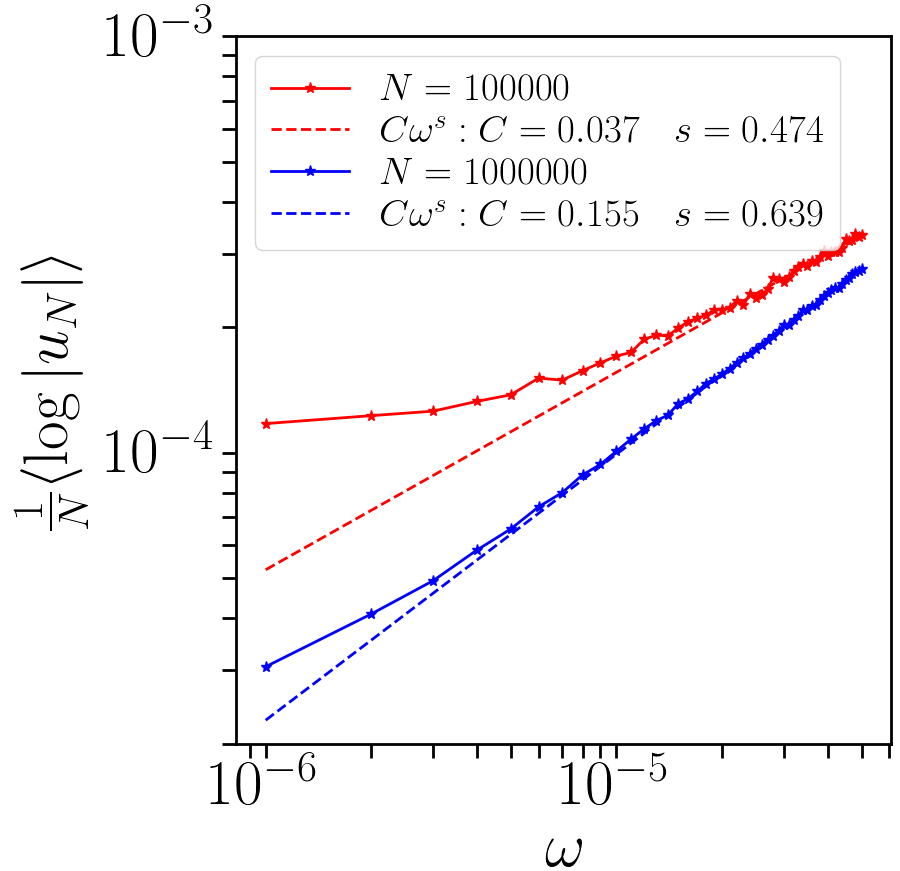}
}
\subfigure[$\expval{B}<0$]{
	\includegraphics[width=5.1cm,height=5.1cm]{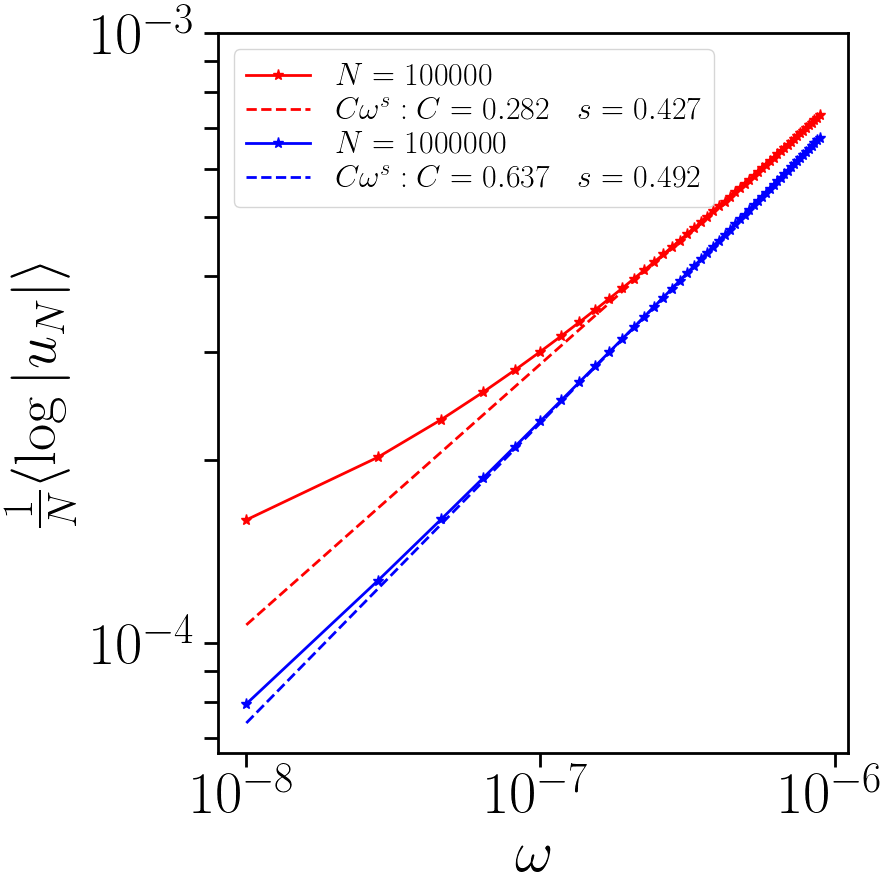}}	
\caption{Variation of numerically calculated Lyapunov exponent, $\lambda=\frac{1}{N}\langle\log \vert u_N \vert \rangle$, with $\omega$. $\langle\log  \vert u_N \vert\rangle$ denotes average of $\log \vert u_N \vert$ over 100 realizations of the random magnetic field.  For (a), (b) and (c) the magnetic fields were chosen randomly from the intervals $(0,1)$, $(-1,1)$ and $(-1,0)$ respectively. The solid line is the data from the simulation while the dashed line is a power law fit, $C\omega^s$, to the data with $C$ and $s$ as fitting parameters. The obtained values of the fitting parameters  agree appreciably with the theoretical values.}
\label{lyp_plts}
\end{figure}

\begin{table}
	\begin{center}
		\begin{tabular}{ |c|c|c|c|c| } 
			\hline
			Case & Range of $B_n$ & $s: \lambda(\omega)\sim C \omega^s$ & $C$ & $C_{\rm{theoretical}}$ \\
			\hline
			\multirow{3}{8em}{$~~\expval{B}>0$\\$\lambda(\omega)\sim \frac{\sigma^2}{8\expval{B}} \omega$} & $(0,0.25)$ & 0.986&0.0045&0.0052 \\ 
			& $(0,0.5)$ & 0.999&0.0102&0.0104 \\ 
			& $(0,0.75)$ &1.0005 &0.0156&0.0156 \\ 
			\hline
			\hline
			\multirow{3}{10em}{$~~~\expval{B}<0$\\$\lambda(\omega)\sim \sqrt{|\langle B\rangle|} \omega^{1/2}$} & $(-0.25,0)$ & 0.492&0.315&0.353 \\ 
			& $(-0.5,0)$ & 0.492&0.444& 0.5\\ 
			& $(-0.75,0)$ &0.491& 0.532&0.612 \\ 
			\hline
			\hline
			\multirow{3}{8em}{$~\expval{B}=0$\\$\lambda(\omega)=\hat\lambda(\sigma)\omega^{2/3}$} & $(-0.25,0.25)$ & 0.658&0.073&0.079 \\ 
			& $(-0.5,0.5)$ & 0.658&0.115&0.127 \\ 
			& $(-0.75,0.75)$ & 0.649&0.136&0.167 \\ 
			\hline
		\end{tabular}
	\end{center}
	\caption{Comparison of analytical prefactor for the three cases with the numerical prefactor. For this table, $N=10^7$.}
		\label{table:prefactorandexponent}
\end{table}
 We numerically calculate the Lyapunov exponents by using Eq.~(\ref{f_N}) to generate $u_N$ for 100 realizations of the random magnetic field. The Lyapunov exponent would then be given by $\lambda=\frac{1}{N}\langle \log \vert u_N \vert \rangle$, where $N$ is the number of oscillators. We plot in Fig.~(\ref{lyp_plts}), the numerical data thus obtained for different $\omega$ and the power law fit, $C\omega^s$, for the data with $C$ and $s$ as fitting parameters. We see that the values of $s$ obtained for the three casses, $\langle B \rangle >0,~\langle B \rangle <0 $ and $\expval{B}=0$, agree reasonably well with the theoretically expected values. The prefactor, $C$, obtained for the three cases also seems to agree with the expected values from theory, see Table \ref{table:prefactorandexponent}. 
 
We  now have the behaviour of the Lyapunov exponents at small $\omega$ for Eq.~(\ref{f_N}) and we found this to be different depending on the expectation value of the random magnetic field. The transmission is determined by $f_N^+$ as well as $f_N^-$ and these two have different Lyapunov exponents for $\expval{B}\neq 0$, therefore the larger of the two exponents will dominate in the transmission. This is the Lyapunov exponent for $f_N^-$ for $\expval{B}>0$, while  for $\expval{B}=0$, $f_N^+$ and $f_N^-$ have the same Lyapunov exponent. In the next section, we determine the size dependence of the current using these results for the Lyapunov exponents.


\section{Size dependence of the current}
\label{sec:size_current}
 We now have the small $\omega$ behaviour of $\lambda(\omega)$ for the transmission. We found this to be different for $\expval{B}\neq 0$ and $\expval{B}=0$, so we expect different power laws  for the current for the two cases. The boundary conditions will also play a role in the power law via the small $\omega$ behaviour of $\mathcal{T}_\infty(\omega)$. We therefore take the cases $\expval{B}\neq0$ and $\expval{B}=0$ separately for the two boundary conditions. 
 \subsubsection*{Fixed boundary conditions:} 
 \begin{itemize}
 \item[(a)] For $\expval{B}\neq0$,  $\mathcal{T}_\infty(\omega)\sim \omega^{3/2}$ and $\lambda(\omega)\sim \omega$. Therefore using these in Eq.~\eqref{currexp1} we have $\langle {J}_N \rangle\sim 1/N^{5/2}$. 
\item[(b)] For $\expval{B}=0$, $\mathcal{T}_\infty(\omega)\sim \omega^{2}$ and $\lambda(\omega)\sim \omega^{2/3}$ which gives $\langle {J}_N \rangle\sim 1/N^{9/2}$.
\end{itemize}

\begin{table}
	\centering
	\label{table}
	\resizebox{\textwidth}{!}{
		\begin{tabular}{ |c|c|c|c|c| }
			\hline
			Boundary Conditions & Average magnetic field & $\mathcal{T}_\infty(\omega)$ & $\lambda(\omega)$ & Power law for the current $\langle {J}_N \rangle$ \\ 
			\hline
			Fixed & $\expval{B}\neq0$ & $\sim \omega^{3/2}$ & $\sim \omega$ & $\sim 1/N^{5/2}$\\
			\hline
			Fixed & $\expval{B}=0$ & $\sim \omega^{2}$& $\sim \omega^{2/3}$ & $\sim 1/N^{9/2}$\\
			\hline
			Free & $\expval{B}\neq0$ & $\sim \omega^{1/2}$& $\sim \omega$ & $\sim 1/N^{3/2}$\\
			\hline
			Free & $\expval{B}=0$ & $\sim \omega^{0}$& $\sim \omega^{2/3}$ & $\sim 1/N^{3/2}$\\
			\hline
		\end{tabular}
	}
	\caption{Power law for the current for different boundary conditions and average magnetic fields.}
\end{table}

\begin{figure}
	\centering
	\subfigure[$\expval{B}\neq0$]{
		\includegraphics[width=7.2cm,height=7cm]{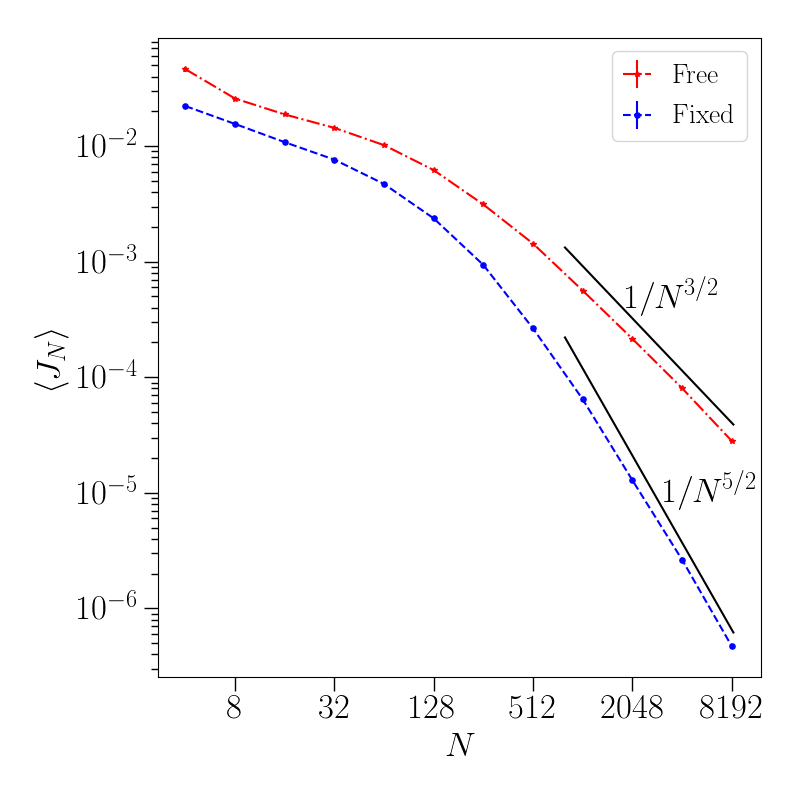}
	}
	\subfigure[ $\expval{B}=0$]{
		\includegraphics[width=7.2cm,height=7cm]{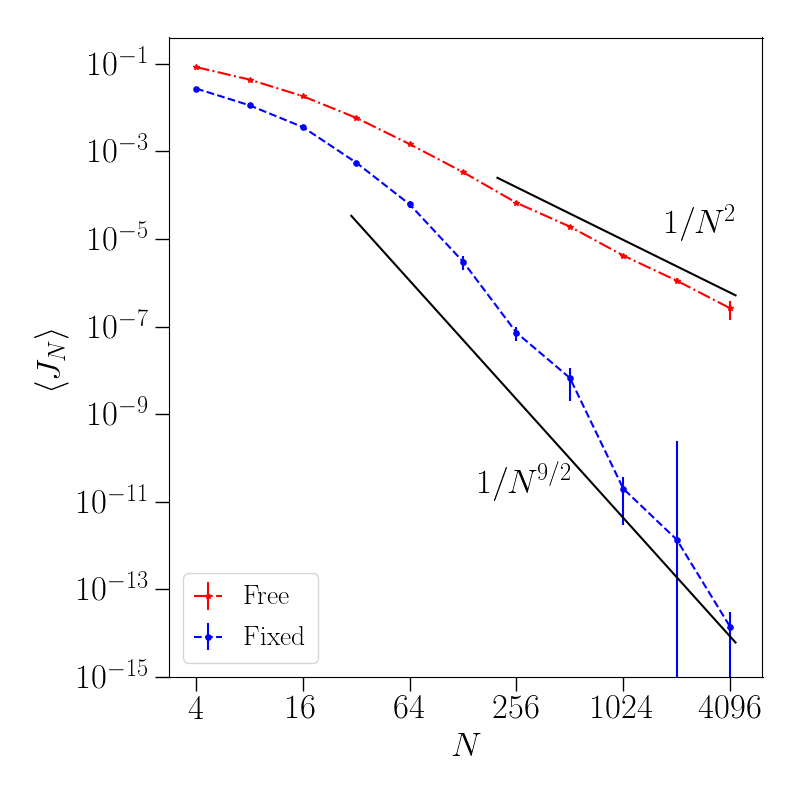}
	}
	\caption{ Numerically obtained power laws for the average current, averaged over $100$ realizations of the disorder, with fixed and free boundary conditions. For $\expval{B}>0$, $B_n$ is chosen from $(1,3)$ while for $\expval{B}=0$, $B_n$ is chosen from $(-2,2)$. }
	\label{curr_plts}
\end{figure}

\subsubsection*{Free  boundary conditions:} 
\begin{itemize}
	\item[(a)] For $\expval{B}\neq0$,  $\mathcal{T}_\infty(\omega)\sim \omega^{1/2}$ and $\lambda(\omega)\sim \omega$ which gives $\langle {J}_N \rangle\sim 1/N^{3/2}$\ . 
	\item[(b)] For $\expval{B}=0$, $\mathcal{T}_\infty(\omega)\sim \omega^{0}$ and $\lambda(\omega)\sim \omega^{2/3}$ which gives $\langle {J}_N \rangle\sim 1/N^{3/2}$ \ .
\end{itemize}
The results are summarized in Table 1. Fig~(\ref{curr_plts}) shows the numerically obtained power laws for  $\expval{B}\neq 0$ and $\expval{B}= 0$. Numerically, the power laws are obtained by calculating $\mathcal{T}_N(\omega)$ for different $\omega$  and then performing the integration numerically. We expect to see the power law behaviour at some large enough $N$. We see a reasonable agreement with the theoretically expected power laws  except for the case with $\expval{B}= 0$ and free BC, where we get $\langle {J}_N \rangle\sim 1/N^{2}$ instead of the expected $\langle {J}_N \rangle\sim 1/N^{3/2}$ .  

 The case with $\expval{B}=0$ seems to be  quite subtle  because of the following reasons:
\begin{itemize}
\item The assumption that $\mathcal{T}_\infty(\omega)$ may be replaced by the transmission for the uniform case for small $\omega$ does not hold good for $\expval{B}=0$ case. This can be clearly seen from  Fig.~(\ref{taucmp_plts}), where we show a comparison of the transmission for small $\omega$ for $\expval{B}\neq 0$ and $\expval{B}=0$ with their respective uniform cases. While $\expval{B}\neq 0$ shows a clear agreement with the corresponding  uniform case, $\expval{B}=0$ case shows a clear disagreement. It is not clear how to estimate $\mathcal{T}_\infty(\omega)$ for this case.
\begin{figure}
	\centering
	\subfigure[$\expval{B}\neq0$]{
		\includegraphics[width=7.9cm,height=6.5cm]{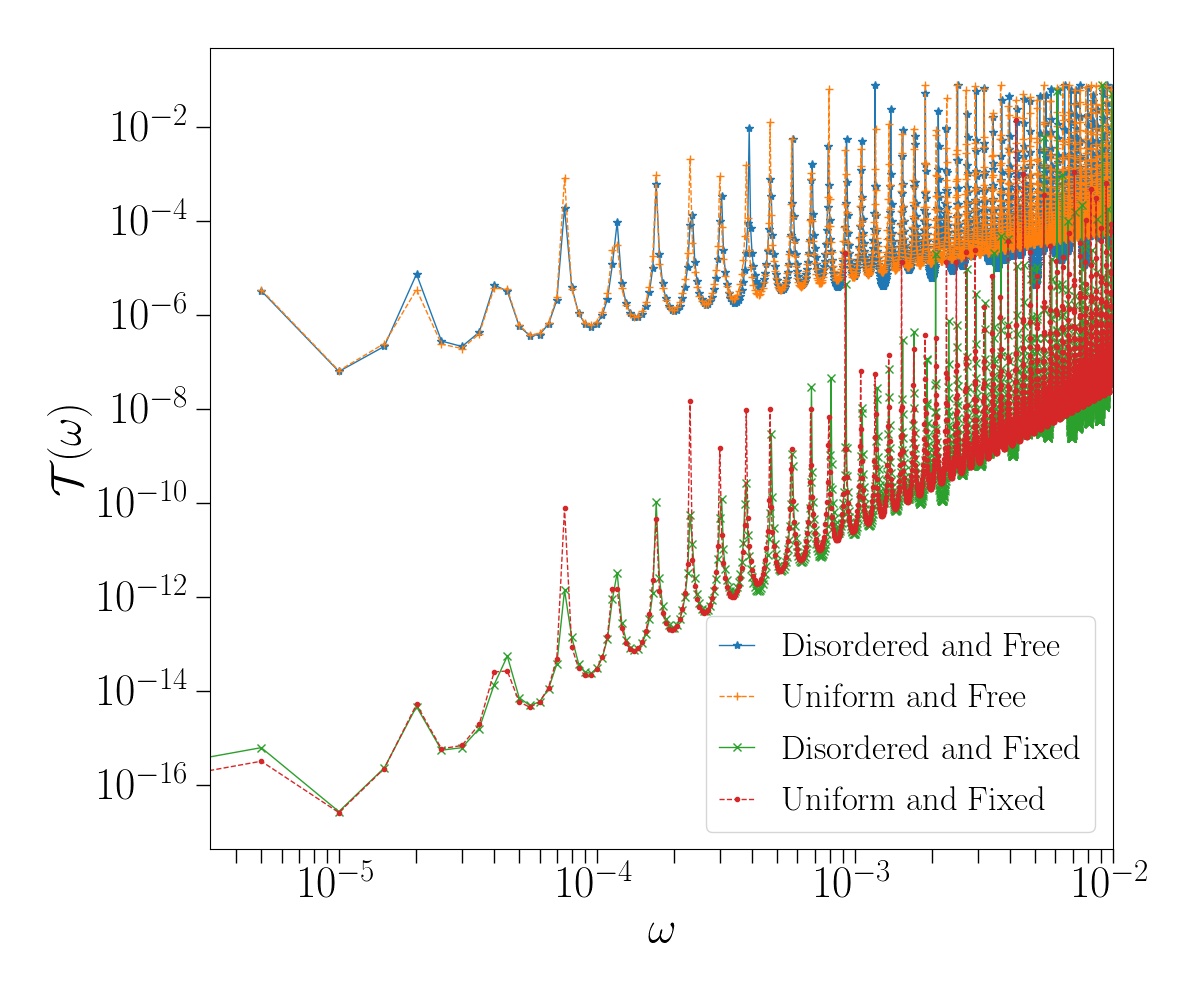}
	}
	\subfigure[ $\expval{B}=0$]{
		\includegraphics[width=7.9cm,height=6.5cm]{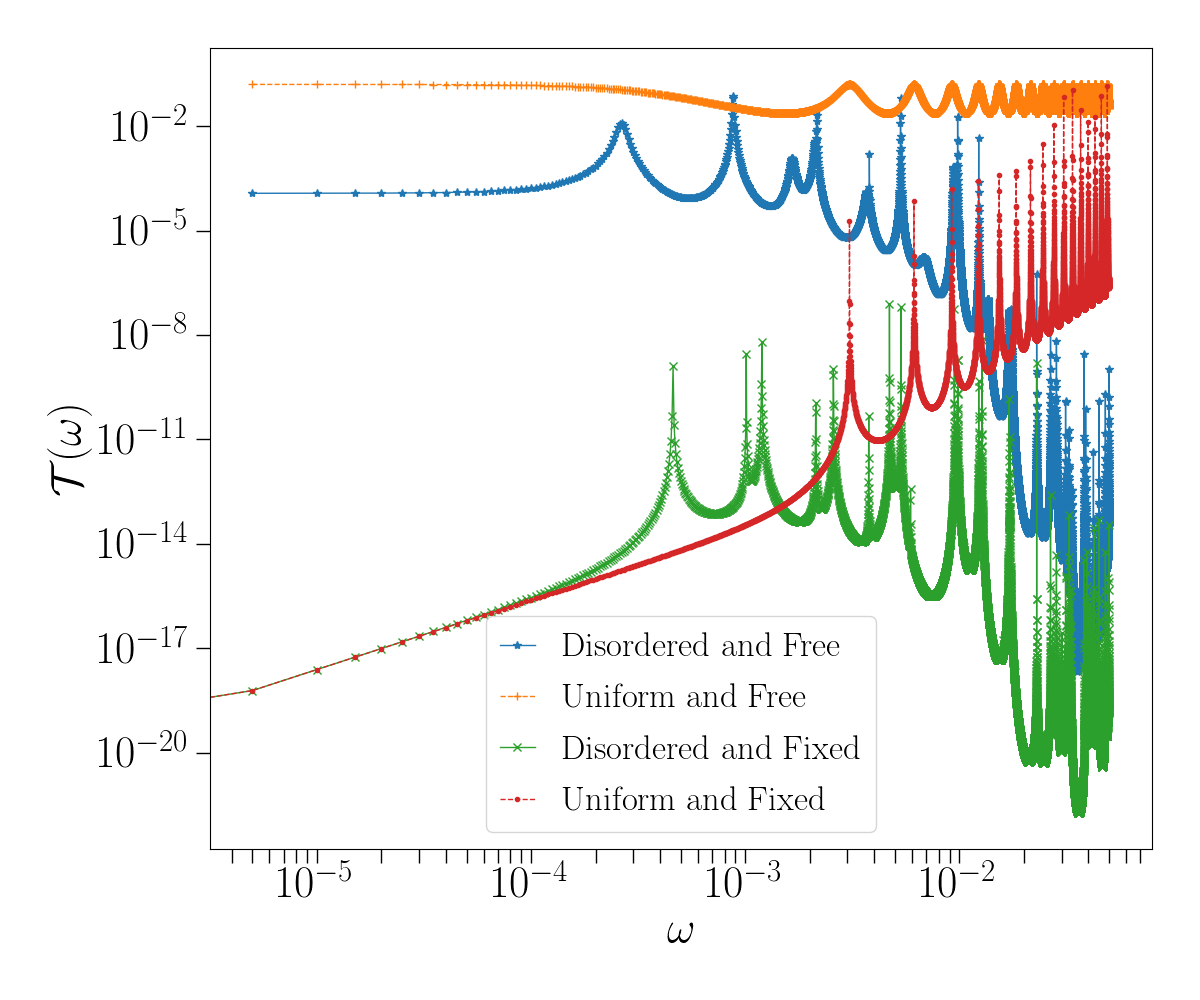}
	}
	\caption{ Comparison for the transmission for disordered and uniform cases for the two boundary conditions. For $\expval{B}\neq 0$, $B_n$ is chosen from $(1,3)$ while for $\expval{B}=0$, $B_n$ is chosen from $(-1,1)$. These are compared with the transmission for the uniform cases with $B_n=\expval{B}$ respectively. }
	\label{taucmp_plts}
\end{figure}

\item\begin{figure}
	\centering
	\subfigure[$\expval{B}\neq0$]{
		\includegraphics[width=8cm,height=6.5cm]{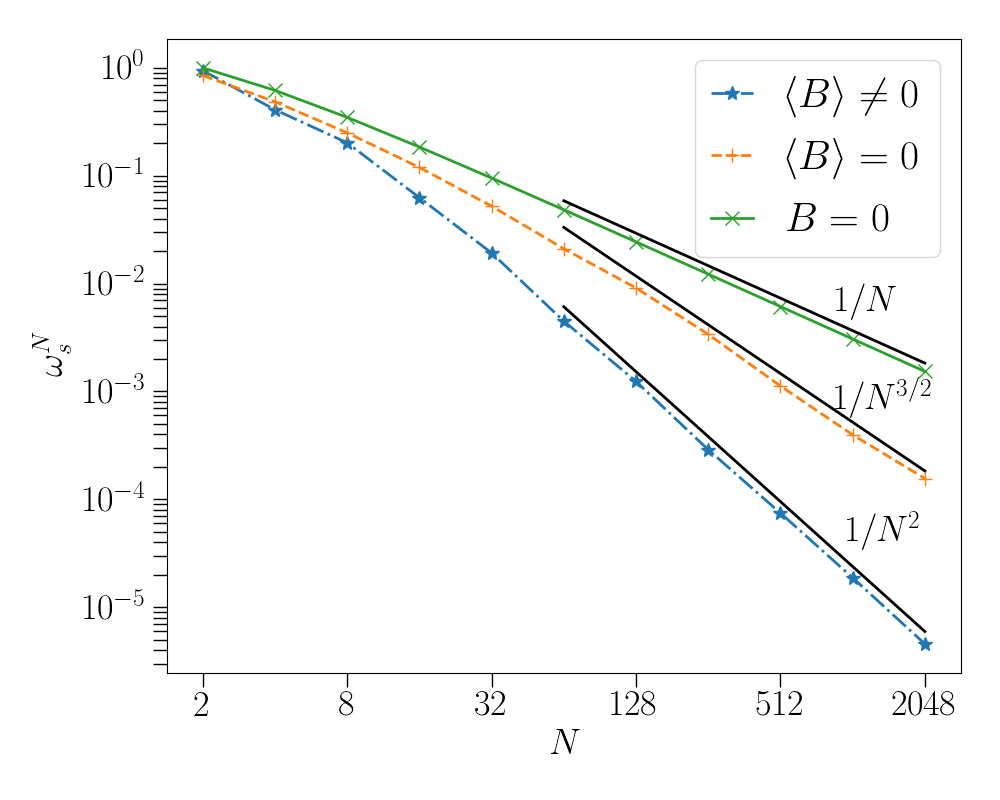}
	}
	\caption{ Scaling of lowest allowed normal mode, $\omega_{s}^N$ with the system size, $N$. For $\expval{B}\neq 0$, $B_n$ is chosen from $(1,3)$ while for $\expval{B}=0$, $B_n$ is chosen from $(-1,1)$.  The $B=0$ plot corresponds to the ordered chain (the ordered case $B\neq 0$ is not shown and has the scaling $N^{-2}$).}
	\label{levalplt}
\end{figure}

 Interestingly we note  that the transmission coefficient has peaks at much lower frequencies than the ordered case. These peaks correspond to the normal modes of the isolated chain and it is then of interest to study the system size dependence of the lowest allowed normal mode frequency,  $\omega_s^N$, for the disordered chains with $\expval{B}\neq 0$ and $\expval{B}= 0$, and the ordered case with $B=0$. In Fig.~(\ref{levalplt}) we show the scaling of $\omega_s^N$ with $N$.  We see that, for $\expval{B}\neq 0$, $\omega_{\rm max}^N\sim 1/N$ while $\omega_{s}^N\sim 1/N^2$. Thus, for any finite but large $N$, we have  $\omega_{\rm max}^N>\omega_{s}^N$  and there are a suffient number of conducting modes. On the other hand, for $\expval{B}=0$,  both $\omega_{max}^N$ and $\omega_{s}^N$ scale as $1/N^{3/2}$ and this could be the reason why our heuristic approach for current scaling fails for this case.
\end{itemize} 

\section{Conclusion}
\label{sec:concl}
We considered a harmonic chain of charged particles in the presence of random magnetic fields and derived power laws for the current with respect to the system size. The power laws were found to be sensitive to boundary conditions and the expectation value of the magnetic field. This was understood as arising from the different  behaviour of the Lyapunov exponent $\lambda (\omega)$ and $\mathcal{T}_\infty(\omega)$ for small frequency $\omega$.  

Arguing that the small $\omega$ behaviour of ${\mathcal T}_\infty (\omega)$ was the same as that for the ordered chain, we used results obtained in our previous paper~\cite{ordered2021} using the non-equilibrium Green function approach. It was found there that this behaviour depends strongly on the presence or not of the magnetic field but also on the boundary conditions imposed. To estimate the Lyapunov exponent we mapped the discrete time process which determines the Green's functions to the motion of a harmonic oscillator with parametric noise. This not only revealed an interesting connection between the Lyapunov exponents of the two systems but also showed that the Lyapunov exponent have different behaviour for different expectation values of the magnetic field. For $\expval{B}>0$, $\expval{B}=0$ and $\expval{B}<0$ we find that the Lyapunov exponents were of order $\omega, ~\omega^{2/3}$ and $\omega^{1/2}$ respectively. These behaviours of the Lyapunov exponent were  also verified numerically. 
 
Using the results for the $\mathcal{T}_\infty(\omega)$ and $\lambda(\omega)$,  we make analytic predictions of  different system-size dependences  of the current, depending on the expectation value of the magnetic field and the boundary conditions. For free boundary conditions the current decreases as $1/N^{3/2}$ irrespective of the expectation value of the magnetic field. However, for fixed boundary conditions the current decreases as $1/N^{3/2}$ and $1/N^{9/2}$ for $\expval{B}\neq 0$ and $\expval{B}=0$ respectively.  Our direct numerical estimates show disagreement for the case $\expval{B}=0$, and this is especially clear for the case with free boundary conditions.  We discussed possible reasons for the disagreement, amongst which is the intriguing numerical observation of the $1/N^{3/2}$ system-size dependence of the lowest normal mode frequency for the  $\expval{B}=0$ case. The resolution of this issue remains an interesting outstanding problem.

\appendix

\section{Lyapunov exponent for a harmonic oscillator with parametric noise}
\label{sec:app-lyap}

In order to obtain an expansion of $\lambda_z (\varepsilon)$ we follow the strategy  developed by Pardoux et al. in \cite{Pardoux_2d} and by Wihstutz in \cite{Wihstutz_ax}. The first step of the proof is to use the ergodic theorem to obtain an explicit formula (see Eq.~\eqref{lyapunov_theta}) for $\lambda_z (\varepsilon)$ instead of Eq.~\eqref{Lyapunov_def_appendix}. In the second step we perform a perturbation analysis in $\varepsilon$ with this new expression.\\

First we express the solution of the $2$-dimensional SDE $(z_t)_{t\ge 0}$ in terms of a $1$-dimensional SDE. Define $(\theta_t)_{t \geq 0}$ to be the solution of 
\begin{equation}
 \dot{\theta}_t  = h_0(\theta_t)  +  \frac{1}{2} \varepsilon^2 \partial_\theta h_1 (\theta_t) h_1(\theta_t) + \varepsilon h_1 (\theta_t) \xi_t \ , 
\label{eq:theta}
\end{equation}
with 
\begin{equation}
\label{def_h0_h1}
h_0(\theta) = \sin^2(\theta)(c-1) - c \quad \text{and} \quad h_1(\theta) = -\sigma\cos^2(\theta)\ .
\end{equation}
One can check that
\begin{align*}
 {z_t}= R_t \ (\cos(\theta_t) , \sin(\theta_t) )^\top
\end{align*}
where
\begin{equation}
\label{SDE_NormX}
 R_t = \Vert z_0 \Vert \exp \left( \int_0^t \left[ q_0 \left( \theta_\tau \right) + \varepsilon^2 r \left( \theta_\tau \right) \right] d\tau - \varepsilon \int_0^t q_1 \left( \theta_\tau \right) \xi_\tau d\tau \right)  ,
\end{equation}
with
\begin{align}
q_0(\theta) &=  (1-c) \cos(\theta) \sin(\theta) \ , \quad q_1(\theta) = \sigma^2\cos(\theta) \sin(\theta) \ ,\label{def_q0}\\
r(\theta) &= \frac{\sigma^2\cos^2 (\theta)}{2} \left[ 2\cos^2(\theta) - 1 \right] \ . 
\label{def_r}
\end{align}
Observe that $\Vert z_t  \Vert =R_t$. Moreover, since in Eq.~\eqref{eq:theta} the noise is vanishing exactly at the points $\theta^*_k= (2k+1)\pi/2$, $k\in \mathbb Z$, and that the drift in Eq.~\eqref{eq:theta} at $\theta_k^*$ is equal to $-1$, we see that starting from $\theta_0 \in [\theta_{k-1}^*, \theta_{k}^*)$ the process $(\theta_t)_{t\ge 0}$ will pass successively in the intervals $\theta_0 \in [\theta_\ell^*, \theta_{\ell+1}^*)$ for $\ell \le k-1$ without coming back to an interval previously visited. This defines a sequence of random times $t_\ell =\inf\{t\ge 0\; ; \; \theta_t  \in [\theta_{k-\ell-1}^*, \theta_{k-\ell}^*) \}$ for $\ell \ge 0$ with $t_0=0$.  The process is thus clearly not ergodic. A simple way to restore this ergodicity (that will be needed later) is to consider the process $({\tilde \theta}_t)_{t\ge 0}$,  living in $[-\pi/2 , \pi/2)$, and defined by $\tilde \theta_t = \theta_t +(k-\ell)\pi$ for $t\in [t_\ell, t_{\ell +1})$. The process ${\tilde \theta}_t$ satisfies the same stochastic differential equation as $(\theta_t)_{t\ge 0}$ but when it reaches $-\pi/2$ it is immediately reseted to $\pi/2$. Equivalently $({\tilde \theta}_t)_{t\ge 0}$ is solution of Eq.~\eqref{eq:theta} but seen as a SDE on the torus $[-\pi/2, \pi/2)$ where the two end points of the interval have been identified. The process $({\tilde \theta}_t)_{t\ge 0}$ has now the nice property to be ergodic. We denote by $\rho_\varepsilon( \theta) d\theta$ its invariant measure which is  computed below. Observe moreover that Eq.~\eqref{SDE_NormX} still holds by replacing $\theta$ by $\tilde \theta$ because the functions $q_0, q_1, r$ are $\pi$-periodic. In order to keep notation simple we denote in the sequel the process $\tilde \theta$ by $\theta$.\\

\medskip
  
By definition \eqref{Lyapunov_def_appendix} of Lyapunov exponent and Eq.~\eqref{SDE_NormX} we get that
\begin{eqnarray*}
\lambda_z (\varepsilon) & = & \lim\limits_{t \rightarrow \infty} \frac{1}{t} \left\langle \int_0^t \left[q_0 \left( \theta_\tau \right) + \varepsilon^2 r \left( \theta_\tau \right) \right] d\tau + \varepsilon \int_0^t q_1 \left( \theta_\tau \right) \xi_\tau d\tau \right\rangle  \\
& = &  \lim\limits_{t \rightarrow \infty} \frac{1}{t} \left\langle \int_0^t \left[q_0 \left( \theta_\tau \right) + \varepsilon^2 r \left( \theta_\tau \right) \right] d\tau \right\rangle, 
\end{eqnarray*}
since $\langle \int_0^t q_1 \left( \theta_\tau \right) \xi_\tau d\tau \rangle =0$. Then by using the ergodic theorem we obtain 
\begin{equation}
\label{lyapunov_theta}
\lambda_z (\varepsilon) = \int_{-\pi/2}^{\pi/2} \left[ q_0(\theta) + \varepsilon^2 r(\theta) \right] \rho_\varepsilon(\theta) d\theta \ .
\end{equation}
The expansion in $\varepsilon$ for $\lambda_z(\varepsilon)$ can then be obtained from the expansion of $\rho_\varepsilon$.\\

\medskip

Before doing this we prove Eq.~\eqref{eq:lambdaxz}, i.e. that the process  $(z_t)_{t \geq 0} = \left( (u_t, v_t)^\top\right)_{ t \geq 0}$ and the process $(u_t)_{t \geq 0}$ have the same Lyapunov exponent. By definition we have
\begin{equation}
\label{proof_lyapunov_exponent_z_t_x_t}
\lim\limits_{ t \rightarrow \infty} \frac{1}{t} \left \langle \log\vert u_t \vert  \right\rangle = \lim\limits_{ t \rightarrow \infty} \frac{1}{t} \left\langle \log\Vert z_t \Vert  \right\rangle + \lim\limits_{ t \rightarrow \infty} \frac{1}{t} \left\langle \log \vert \cos(\theta_t) \vert  \right\rangle.
\end{equation}
Since $(\theta_t)$ is an ergodic process we obtain that 
\[ \lim\limits_{ t \rightarrow \infty} \frac{1}{t} \left\langle \log \vert \cos(\theta_t) \vert  \right\rangle= \lim\limits_{ t \rightarrow \infty} \frac{1}{t} \int_{-\pi/2}^{\pi/2} \rho_\varepsilon(\theta) \log\left(  \vert \cos(\theta) \vert \right) d\theta = 0 \ .  \]
This proves the claim.

\medskip

Let us now compute $\rho_\varepsilon$ which is the solution of the stationary Fokker-Planck equation
\begin{equation}
\label{eq:FP}
\partial_\theta \left[ \tfrac{\varepsilon^2 }{2} \partial_\theta (h_1^2 \rho_\varepsilon) -(h_0 +\tfrac{\varepsilon^2}{2} h_1 \partial_\theta h_1  ) \rho_\varepsilon\right]=0 \ .
\end{equation}
If we look for a solution such that $\tfrac{\varepsilon^2}{2} \partial_\theta (h_1^2 \rho_\varepsilon) -(h_0 +\tfrac{\varepsilon^2}{2} h_1 \partial_\theta h_1  ) \rho_\varepsilon=0$ we get $\rho_\varepsilon (\theta)  \propto \cos^{-2} (\theta) e^{-\frac{2 \varepsilon^{-2}}{3 \sigma^2} \tan^3 (\theta) -\frac{2c \varepsilon^{-2}}{ \sigma^2}\tan(\theta)}$ which is not normalisable. Hence we have to look for a normalisable solution such that  $\tfrac{\varepsilon^2}{2} \partial_\theta (h_1^2 \rho_\varepsilon) -(h_0 +\tfrac{\varepsilon^2}{2} h_1 \partial_\theta h_1  ) \rho_\varepsilon=A$ for some constant $A$. We get then that
\begin{equation*}
\rho_\varepsilon (\theta) =Z_\varepsilon^{-1} v_\varepsilon (\theta)\cos^{-2} (\theta) \int_{- \infty}^{\tan(\theta)}  \  \exp\left( \frac{2 \varepsilon^{-2}}{3 \sigma^2} u^3  +\frac{2c \varepsilon^{-2}}{ \sigma^2}u \right) du
\end{equation*}
with
\begin{equation*}
v_\varepsilon (\theta)= \exp\left\{ -\frac{2 \varepsilon^{-2}}{3 \sigma^4} \tan^3 (\theta) -\frac{2c \varepsilon^{-2}}{ \sigma^4}\tan(\theta) \right\}
\end{equation*}
and $Z_\varepsilon$ the partition function making $\rho_\varepsilon$ a probability. Injecting this in Eq.~\eqref{lyapunov_theta} we may derive the results claimed by a careful saddle point analysis. We prefer instead to rely on a more heuristic analysis to bypass boring computations.\\

It is natural to expect that as $\varepsilon \to 0$ the stationary measure $\rho_\varepsilon (\theta) d\theta$ will converge to the one of ${\dot \theta}_t =h_0 (\theta_t)$ (i.e. Eq.~\eqref{eq:theta} with $\varepsilon=0$). However as we will see this deterministic dynamical system has different behaviours depending on the value of $c$ and that in some cases we have also to compute the next order corrections.\\

\medskip 

If $c > 0$, the deterministic dynamical system has a unique invariant state $\rho_0 (\theta) d\theta$ with  $\rho_0 (\theta) = -\frac{\sqrt{c}}{\pi} h_0^{-1}(\theta)$ because $h_0$ never vanishes on $[-\pi/2 , \pi/2)$. Hence $\rho_\varepsilon \to \rho_0$ as $\varepsilon \to 0$. However, since  $\int_0^\pi q_0(\theta) \rho_0 (\theta) d\theta = 0$, we have to expand $\rho_\varepsilon$ at order $\varepsilon^2$ to obtain the behavior of $\lambda_\varepsilon$ in Eq.~\eqref{lyapunov_theta}. Let us assume that $\rho_\varepsilon =\rho_0 +\varepsilon^2 \delta\rho_0 + o(\varepsilon^2)$, inject this in Eq.~\eqref{eq:FP} and identify the powers in $\varepsilon$. We obtain that 
\begin{equation*}
\partial_\theta [h_0 \ (\delta \rho_0)]= \tfrac{1}{2}\partial_\theta \left[  \partial_\theta (h_1^2 \rho_0) -(h_1 \partial_\theta h_1  ) \rho_0\right]
\end{equation*}
which implies, since $\int_{-\pi/2}^{\pi/2} (\delta \rho_0) (\theta) d\theta =0$ that
\begin{equation*}
\delta\rho_0 = \tfrac{A}{h_0}+\tfrac{1}{2h_0}\left[  \partial_\theta (h_1^2 \rho_0) -(h_1 \partial_\theta h_1  ) \rho_0\right].
\end{equation*}
We deduce that 
\begin{equation*}
\delta\rho_0 = \frac{A}{h_0} +\frac{\sigma^2\sqrt{c} }{\pi}\left(\frac{\sin(\theta)\cos^3(\theta) }{h_0^2}+ (c-1)\left(  \frac{\cos^5(\theta) \sin(\theta)}{h_0^3(\theta)}  \right) \right).
\end{equation*}
Since $\int_{-\pi/2}^{\pi/2} (\delta \rho_0) (\theta) d\theta =0$ we obtain $A=0$ and
\begin{equation*}
\delta\rho_0= \frac{\sigma^2 \sqrt{c} }{\pi}\left(\frac{\sin(\theta)\cos^3(\theta) }{h_0^2}+ (c-1)\left(  \frac{\cos^5(\theta) \sin(\theta)}{h_0^3(\theta)}  \right) \right).
\end{equation*}
Hence we get that
\begin{equation*}
\lambda_z (\varepsilon)  = \varepsilon^2 \int_{- \pi/2}^{\pi /2} \left(r(\theta) \rho_0(\theta)+q_0(\theta) \delta \rho_0(\theta) \right) d\theta + o(\varepsilon^2) \ .
\end{equation*}
By the change of variable $x=\tan(\theta)$ we get
\begin{align*}
\int_{- \pi/2}^{\pi /2} r(\theta) \rho_0(\theta)d\theta &= \frac{\sigma^2\sqrt{c}}{2\pi } \int_{-\infty}^\infty \frac{x^2-1}{(1+x^2)(x^2+c)} dx = \frac{\sigma^2}{2(\sqrt{c}+1)^2} .  \\ 
\int_{- \pi/2}^{\pi /2} q_0(\theta) \delta \rho_0(\theta) d\theta &= \sigma^2\left(\frac{(4 \sqrt{c} +1)(c-1)^2}{8(\sqrt{c}+1)^4 c} + \frac{1-c}{2(\sqrt{c}+1)^3}\right). \\ 
\end{align*}
Hence we finally get
\[ \lambda_z (\varepsilon) = \varepsilon^2 \frac{\sigma^2}{8c} + o(\varepsilon^2) . \]
This proves case (ii).
\medskip

If $c<0$ then $\tfrac{c}{c-1}\in (0,1)$ and the function $h_0$ vanishes on $[-\pi/2, \pi/2)$ if and only if $\theta \in [-\pi/2, \pi/2)$ is solution of 
$$\sin^2(\theta) = \frac{c}{c-1}\ .$$
There are two solutions $\theta^*>0$ and $-\theta^*<0$. The deterministic dynamical has two extremal  invariant probability measures given by $\delta_{\pm \theta^\star}$. Since $h^\prime_0(\theta^*)<0<h^\prime_0 (-\theta^*)$, $\delta_{-\theta^*}$ is unstable while $\delta_{\theta_*}$ is stable. By introducing noise in this dynamical system the stable stationary state is selected when the intensity of the noise is sent to zero afterwards, i.e. $\rho_\varepsilon (\theta) d\theta \to \delta_{\theta^*}$. We conclude that 
\[ \lim_{\varepsilon \to 0}\lambda_z (\varepsilon) =  q_0(\theta^*) = \sqrt{\vert c \vert } \ .  \]
This proves case (iii).

\medskip 

The case $c=0$ is more delicate. Since $h_0( \cdot) = -\sin^2(\cdot)$, the unique invariant measure for the deterministic dynamical system is $\delta_0$ (stable) and we expect that $\rho_\varepsilon (\theta) d\theta \to \delta_0$ as $\varepsilon \to 0$. Observe however that $q_0 (0) = 0$ so that we have to find the first correction to the approximation of $\rho_\varepsilon$ to $\delta_0$. Due to the singularity of the Dirac mass we cannot perform an expansion analysis in $\varepsilon$. Hence we will use another argument to get item (i). Consider the following linear transformation $T_\varepsilon = \left( \begin{smallmatrix} \varepsilon^{2/3} &0 \\ 0 & 1  \end{smallmatrix}\right)$ which is such that $\varepsilon^{2/3} \Vert z \Vert \leq \Vert T_\varepsilon z \Vert \leq \Vert z \Vert$ for any $z\in \mathbb R^2$ and $\varepsilon \le 1$. This implies that $(z_t)_{t\ge 0}$ and $(T_{\varepsilon}z_t)_{t\ge 0}$ have the same Lyapunov exponent. Expressing as we did before 
\begin{equation*}
{\hat z}_t :=T_{\varepsilon} z_t = \Vert {\hat z}_t \Vert (\cos {\hat \theta}_t,  \sin {\hat \theta}_t)^\top
\end{equation*}
we notice that
\[  \dot{\hat \theta}_t = \varepsilon^{2/3}\left( - \sin^2({\hat\theta}_t) -\sigma^2\sin ({\hat\theta}_t) \cos^3({\hat\theta}_t) \right) - \varepsilon^{1/3} \sigma\cos^2({\hat\theta}_t) \xi_t \ , \]
which implies by scaling invariance of the white noise that ${\hat \theta}_t =\alpha_{t\varepsilon^{2/3}}$ where 
\[\dot \alpha_t = \left( - \sin^2(\alpha_t) -\sigma^2\sin (\alpha_t) \cos^3(\alpha_t) \right) -\sigma  \cos^2({\alpha}_t) \xi_t\ . \]
If ${\hat \rho}(\alpha) d\alpha$ is the unique invariant measure for $(\alpha_t)_{t\ge 0}$ we have by a scaling argument that the Lyapunov exponent satisfies 
 \begin{equation*}
\lambda_z ( \varepsilon) = \varepsilon^{2/3} {\hat \lambda} (\sigma) 
\end{equation*}
with 
\begin{equation}
\label{expression_coef_<B>=0}
 \hat{\lambda} (\sigma)= \int_{- \pi/2}^{\pi/2} \left(q_0(\alpha) + r(\alpha) \right) \hat{\rho}(\alpha) d\alpha,
\end{equation}
where $q_0$ and $r$ are defined respectively in Eq.~\eqref{def_q0} and Eq.~\eqref{def_r} with $c=0$. To obtain the value of $\hat \lambda (\sigma)$ it is sufficient to find $\hat\rho$ which is the unique normalisable function of the Fokker-Planck equation associated to the process$(\alpha_t)_{t\ge 0}$, i.e.
\begin{equation*}
{\hat \rho} (\alpha)= {\hat Z}^{-1} \cos^{-2} (\alpha) e^{-\tfrac{2}{3\sigma^2} \tan^3 (\alpha)} \ \int_{-\infty}^{\tan (\alpha)} \exp (\tfrac{2u^3}{3\sigma^2}) du\ ,
\end{equation*}
 where ${\hat Z}$ is the normalisation constant making $\hat\rho$ a probability measure.

\section*{Acknowledgements}
A.D. and J.M.B. acknowledge support of the Department of Atomic Energy, Government of India, under Project No. RTI4001. The work of C.B. and G.C. has been supported by the projects LSD ANR-15-CE40- 0020-01 of the French National Research Agency (ANR), by the European Research Council (ERC) under the European Unions Horizon 2020 research and innovative programme (grant agreement No 715734) and the French-Indian UCA project `Large deviations and scaling limits theory for non-equilibrium systems'.

\section*{References}
\bibliographystyle{ieeetr}
\bibliography{biblio}

\end{document}